\documentclass[11pt,a4paper]{article}
\usepackage[T1]{fontenc}
\usepackage[utf8]{inputenc}
\usepackage{graphicx}
\usepackage{amsmath,amssymb}
\usepackage{cite}
\usepackage[colorlinks=true,linkcolor=blue,citecolor=blue,urlcolor=blue]{hyperref}
\usepackage{orcidlink}
\usepackage{QCDEX2026}
\newcommand{\PaperTitle}{Longitudinal Parton Correlations in MRW-Type Unintegrated Double Parton Distributions}

\newcommand{\PaperAbstract}{
Double parton scattering offers a way to investigate correlations between partons inside the proton that are absent from the usual product approximation for double parton distributions. In this contribution, we discuss how these correlations enter unintegrated double parton distributions when transverse momentum is generated using MRW-type prescriptions. Starting from non-factorized GS09 distributions evolved to unequal scales, we examine the double LO-MRW, virtuality-ordered MRW, and modified KMRW constructions, focusing on their transverse-momentum dependence and flavor structure.
}

\begin{document}
\noindent\includegraphics[width=0.4\textwidth]{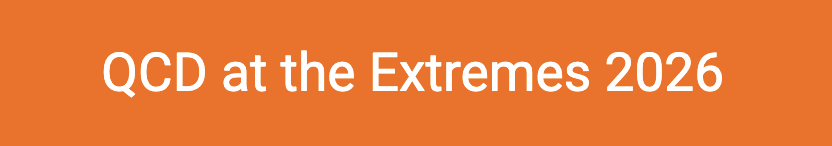}\par
\vspace{0.6em}
\begin{center}
{\LARGE\bfseries \PaperTitle\par}
\vspace{0.8em}
{\large R.~Kord Valeshabadi$^{1,*}$\,\orcidlink{0000-0001-5326-2459},\quad
S.~Rezaie$^1$\,\orcidlink{0000-0002-1852-1619},\quad
and K.~Azizi$^{1,2,3}$\,\orcidlink{0000-0003-3741-2167}\par}
\vspace{0.5em}
{\small
$^1$School of Particles and Accelerators, Institute for Research in Fundamental Sciences (IPM), Tehran, Iran\par
$^2$Department of Physics, University of Tehran, Tehran, Iran\par
$^3$Department of Physics, Dogus University, Istanbul, Turkey\par}
\vspace{0.35em}
{\small $^*$Speaker and corresponding author: \href{mailto:ramin.kord@ipm.ir}{ramin.kord@ipm.ir}\par}
\vspace{0.35em}
{\small\itshape refereed by the conveners of QCDEX2026\par}
\end{center}
\vspace{0.5em}
\begin{abstract}
\PaperAbstract
\end{abstract}

\section{Introduction}

Double parton scattering (DPS) involves two hard partonic interactions in the same hadron--hadron collision. A common approximation treats the two partons extracted from a proton as independent and replaces their joint distribution by a product of ordinary PDFs. This is convenient, but it misses correlations from the proton's finite valence content and from QCD splitting. Those correlations become important when the available longitudinal momentum is limited or when particular flavor combinations are selected \cite{Gaunt:2009re}.

For observables involving transverse-momentum imbalance or azimuthal correlations, it is also useful to keep the transverse momenta of the two incoming partons. This leads to unintegrated double parton distribution functions (UDPDFs),
$\mathcal F_{ij}(x_1,x_2,k_{1t}^2,k_{2t}^2;\mu_1^2,\mu_2^2)$.
We examine how longitudinal correlations are reflected in the transverse-momentum dependence when UDPDFs are constructed using prescriptions inspired by the Martin--Ryskin--Watt (MRW) approach \cite{MRW}. Comparing the correlated distributions with independent-parton reference distributions within each prescription helps to assess these effects separately from differences among the transverse-momentum models. This contribution concentrates on selected numerical results, and the full calculation is described in Ref.~\cite{Valeshabadi:2026ckt}.

\section{Correlated collinear input}

We use the GS09 input \cite{Gaunt:2009re}, which incorporates momentum and valence-number constraints. For comparison, we also use the product approximation
\begin{equation}
	D_{ij}^{\mathrm{fact}}(x_1,x_2;\mu_1^2,\mu_2^2)
	=f_i(x_1,\mu_1^2)f_j(x_2,\mu_2^2)\,
	\Theta(1-x_1-x_2),
	\label{eq:factorized}
\end{equation}
which neglects the dynamical correlations between the two partons. The GS09 input instead contains flavor-dependent correlations. At leading order the equal-scale double DGLAP equation has the schematic form~\cite{Gaunt:2009re,Ceccopieri:2010kg}
\begin{equation}
	\frac{\partial D_{ij}}{\partial\ln\mu^2}
	=\frac{\alpha_s(\mu^2)}{2\pi}
	\bigl[P\otimes_1 D+P\otimes_2 D+S_{ij}[f]\bigr].
	\label{eq:ddglap}
\end{equation}
The first two terms evolve the two partons separately. The source term $S_{ij}[f]$ accounts for a single parent parton splitting into the pair. We keep both the homogeneous and non-homogeneous pieces, $D_{ij}=D_{ij}^{(h)}+D_{ij}^{(nh)}$.

For unequal scales, we first evolve both partons to the lower scale and then evolve only the leg associated with the higher scale \cite{Ceccopieri:2010kg,Gaunt:2009re}. We perform the evolution with \textsc{ChromaPDFEvolver} and interpolate the resulting grids using \textsc{PDFxTMDLib} \cite{Valeshabadi2026,Valeshabadi:2026ckt}. The numerical checks of this evolution are reported in Ref.~\cite{Valeshabadi:2026ckt}.

\section{Generating the transverse momenta}

We compare three prescriptions applied to the same collinear input: double LO-MRW (DLO-MRW), double virtuality-ordered MRW (DVO-MRW), and double modified KMRW (DMKMRW). They differ primarily in how transverse momentum is introduced and at which scales the underlying DPDF is sampled.

\paragraph{Double LO-MRW (DLO-MRW).}
In this approach, the final real-emission step is unfolded on each evolution ladder. The Sudakov factors $T_i$ and $T_j$ encode the probability of no resolvable emission between the transverse-momentum scales and the hard scales. For $k_{1t},k_{2t}\ge\mu_0$, the result has the schematic form
\begin{equation}
	\mathcal F^{\mathrm{DLO}}_{ij}\sim
	T_iT_j\,
	\frac{\alpha_s(k_{1t}^2)}{2\pi k_{1t}^2}
	\frac{\alpha_s(k_{2t}^2)}{2\pi k_{2t}^2}
	\bigl(P^R\otimes_1P^R\otimes_2D\bigr)_{ij}.
	\label{eq:dlo}
\end{equation}
The expression can be evaluated with the full evolved DPDF in this perturbative region. Below $\mu_0$, however, the conventional MRW model needs separate treatments depending on which transverse momentum lies below the starting scale. The non-homogeneous contribution makes this extension particularly awkward \cite{Golec-Biernat:2016vbt,Valeshabadi:2026ckt}. We therefore use DLO-MRW only where both transverse momenta are perturbative.

\paragraph{Double virtuality-ordered MRW (DVO-MRW).}
Here the scale of each last branching is its virtuality rather than its transverse momentum:
\begin{equation}
	K_a^2=\frac{k_{at}^2}{1-z_a},\qquad
	\mu_0^2\le K_a^2\le\mu_a^2 \quad (a=1,2).
	\label{eq:virtuality}
\end{equation}
This model employs LO splitting kernels with the virtuality-ordered scale choice motivated by the MRW approach. The corresponding double distribution takes the schematic form
\begin{align}
	\mathcal F_{ij}^{\mathrm{DVO}}
	&=\sum_{a,b}\int\frac{dz_1}{z_1}\frac{dz_2}{z_2}
	\prod_{r=1}^{2}\!\left[
	\frac{\alpha_s(K_r^2)}{2\pi k_{rt}^2}
	T_r(K_r^2,\mu_r^2)\right]\nonumber\\[-2pt]
	&\quad\times P^R_{a\to i}(z_1)P^R_{b\to j}(z_2)
	D_{ab}\!\left(\frac{x_1}{z_1},\frac{x_2}{z_2};K_1^2,K_2^2\right)
	\Theta_{\mathrm{cuts}}.
	\label{eq:dvo}
\end{align}
Here the cuts enforce angular and virtuality ordering and the two-parton kinematic constraint $x_1/z_1+x_2/z_2\le 1$. It can be used with the full DPDF without the separate low-$k_t$ pieces needed in DLO-MRW. However, integration over the transverse momenta does not generally recover the input collinear DPDF \cite{Valeshabadi:2026ckt}.

\paragraph{Double modified KMRW (DMKMRW).}
In this prescription the full DPDF is evaluated at the hard scales, while the transverse-momentum dependence is supplied by two kernels derived from modified Sudakov factors \cite{Guiot:2022psv}:
\begin{equation}
	\mathcal F_{ij}^{\mathrm{DMKMRW}}
	=D_{ij}(x_1,x_2;\mu_1^2,\mu_2^2)
	\mathcal K_i(k_{1t}^2,\mu_1^2)
	\mathcal K_j(k_{2t}^2,\mu_2^2).
	\label{eq:dmkmrw}
\end{equation}
The essential difference from DLO-MRW and DVO-MRW is that the correlated DPDF is an overall factor, rather than being sampled inside the last-step splitting integrals. The kernels depend on flavor and scales, but not on the correlated two-parton input.

\section{Transverse-momentum dependence and longitudinal correlations}

Our numerical comparison uses the correlated GS09 DPDFs and the corresponding MSTW2008LO single PDFs \cite{MSTW,Gaunt:2009re}. For the factorized reference in the numerical plots, we take the product of the corresponding single UPDFs constructed with each prescription \cite{Valeshabadi:2026ckt}. We consider the $gg$, $ug$, $u\bar u$, and $uu$ channels. To compare each correlated distribution with its independent-parton reference, we form the ratio
\begin{equation}
	R_{ij}^{\mathrm{model}}(x_1,x_2,k_{1t}^2,k_{2t}^2;\mu_1^2,\mu_2^2)
	=\frac{\mathcal F_{ij}^{\mathrm{model,nonfact}}}
	{\mathcal F_{ij}^{\mathrm{model,fact}}}.
	\label{eq:ratio}
\end{equation}
The numerator and denominator use the corresponding transverse-momentum prescription. Hence $R_{ij}=1$ indicates agreement with the independent-parton reference at that point in phase space. Deviations from unity show the effect of departing from this approximation, including the correlations and two-parton phase-space constraints carried by the full DPDF.

For DMKMRW, the transverse kernels cancel between numerator and denominator, leaving
\begin{equation}
	R_{ij}^{\mathrm{DMKMRW}}
	=\frac{D_{ij}(x_1,x_2;\mu_1^2,\mu_2^2)}
	{f_i(x_1,\mu_1^2)f_j(x_2,\mu_2^2)}.
	\label{eq:cancel}
\end{equation}
At fixed longitudinal fractions and scales, the DMKMRW ratio therefore does not vary with either transverse momentum. The situation is different for DLO-MRW and DVO-MRW. In those models the DPDF appears inside the real-emission convolutions, which probe the parent momentum fractions $x_1/z_1$ and $x_2/z_2$ at scales determined by the last branchings. Changing $k_t$ changes which parts of the DPDF contribute most strongly. A longitudinal correlation can consequently alter the $k_t$ shape, not just the magnitude of the distribution.

\begin{figure}[!htbp]
	\centering
	\includegraphics[width=0.45\linewidth]{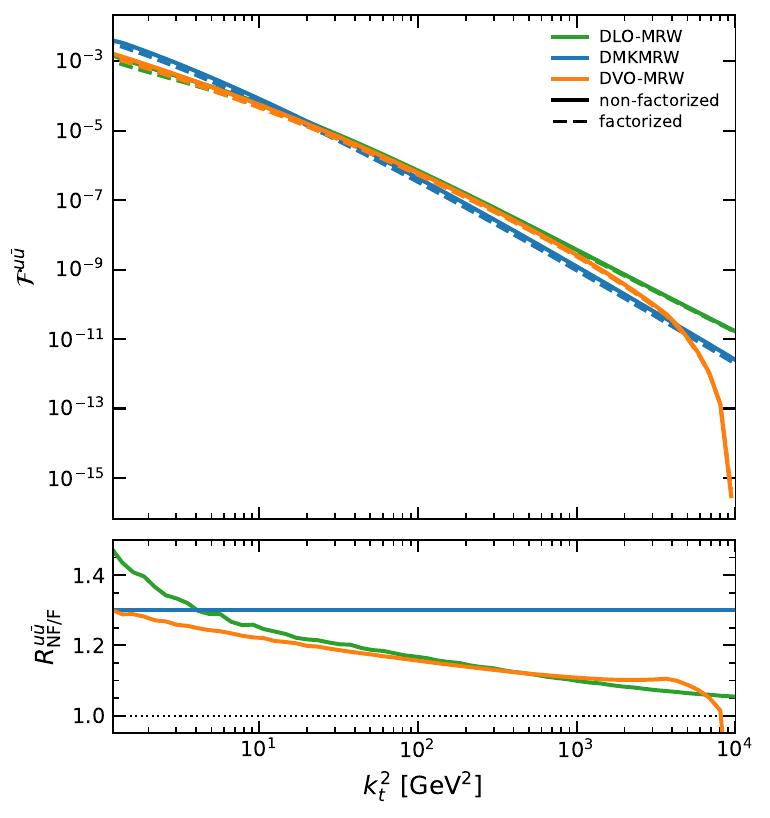}
	\includegraphics[width=0.45\linewidth]{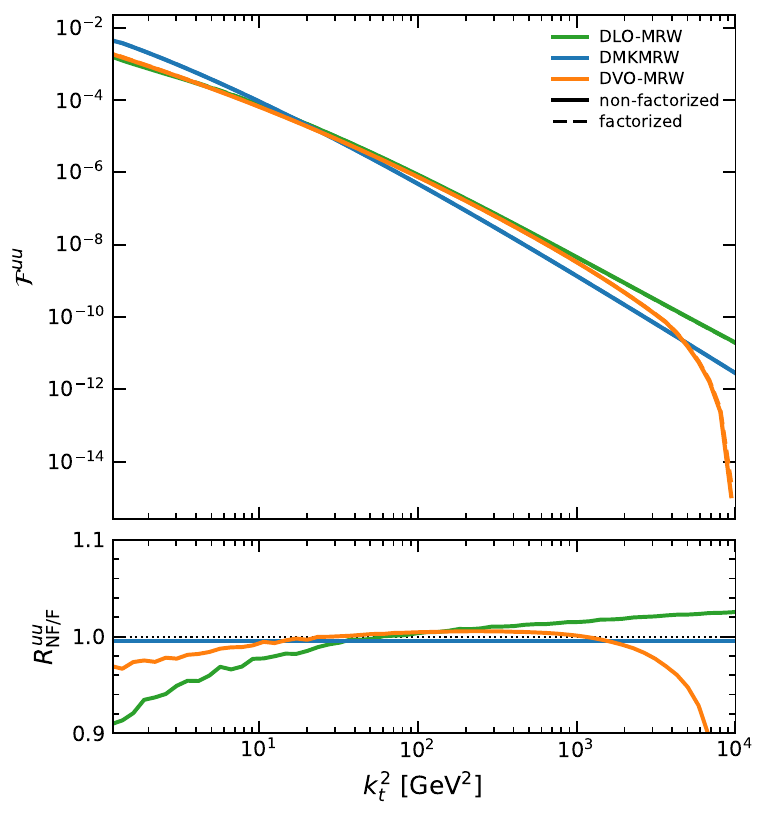}
	\caption{Transverse-momentum dependence of the $u\bar u$ and $uu$ UDPDFs at $x_1=x_2=0.01$, $k_{1t}^2=k_{2t}^2=k_t^2$, and $\mu_1=\mu_2=100\,\mathrm{GeV}$. The upper panels compare DLO-MRW, DVO-MRW, and DMKMRW; solid and dashed curves indicate non-factorized and factorized inputs. The lower panels show the corresponding ratios $R_{ij}$. Adapted from panels (c,d) of Fig.~4 in Ref.~\cite{Valeshabadi:2026ckt} (CC BY 4.0).}
	\label{fig:ktdep}
\end{figure}

Figure~\ref{fig:ktdep} shows that the effect of correlations depends on the flavor channel. In the $uu$ channel, the non-factorized distribution is suppressed relative to the factorized one over much of the range. This is related to the GS09 valence-number correction, where selecting one valence $u$ quark leaves fewer valence $u$ quarks available for the second parton than the independent-parton ansatz would imply. The full $uu$ channel also includes sea-quark contributions. The $u\bar u$ channel behaves differently. Its non-factorized distribution can be enhanced over a substantial part of the plotted range, reflecting the quark--antiquark correlations included in the GS09 input \cite{Gaunt:2009re,Valeshabadi:2026ckt}. The ratio curves also illustrate the distinction between the models. DMKMRW gives a $k_t$-independent correction, whereas the DLO-MRW and DVO-MRW ratios vary with $k_t$. For the $gg$ and $ug$ channels, the deviations from the product approximation are more modest at these small momentum fractions \cite{Valeshabadi:2026ckt}.

The absolute transverse-momentum spectra reveal a further difference. The DVO-MRW distributions fall steeply near the upper end of the range because $K_a^2=k_{at}^2/(1-z_a)\leq\mu_a^2$ restricts the allowed last-branching phase space. At increasing $k_{at}$, the allowed interval in $z_a$ narrows. This decrease is driven by the ordering constraint. The behavior of the ratio $R_{ij}$, by contrast, probes departures from the independent-parton approximation and their flavor dependence.

Finally, the correlation effects are not confined to the small-$x$ example in Fig.~\ref{fig:ktdep}. The last-step convolutions sample the DPDF at $x_1/z_1$ and $x_2/z_2$, subject to $x_1/z_1+x_2/z_2\leq 1$. When larger momentum fractions are selected, the available two-parton phase space becomes more restricted and the differences from the product approximation can be considerably larger, especially in valence-sensitive channels. A wider scan in the $(x,k_t)$ plane for DVO-MRW is given in Fig.~5 of Ref.~\cite{Valeshabadi:2026ckt}. Thus, correlations in the collinear DPDF remain visible after transverse momentum has been generated and, in the last-step prescriptions, acquire a nontrivial dependence on $k_t$.

\section{Conclusions}

Our comparison shows that longitudinal DPDF correlations enter the three UDPDF prescriptions differently. In DMKMRW, the correlated DPDF multiplies transverse kernels, and the non-factorized-to-factorized ratio is independent of $k_t$ at fixed $x_1,x_2$ and scales. In DLO-MRW and DVO-MRW, the DPDF is evaluated inside the last-step branching convolutions, allowing the same longitudinal correlations to change the transverse-momentum distribution. The virtuality bound in DVO-MRW additionally produces a steep decrease near the upper $k_t$ limit.

The contrast between $uu$ and $u\bar u$ illustrates why the flavor structure of the input matters. The valence-number constraints tend to suppress the former relative to the product approximation, while quark--antiquark correlations can enhance the latter.

\section*{Acknowledgements}
This contribution is based on our previously published work in Ref.~\cite{Valeshabadi:2026ckt}. We thank the organizers of \emph{QCD at the Extremes 2026} for the opportunity to discuss these results.

\begingroup
\footnotesize
\bibliographystyle{QCDEX2026}
\bibliography{QCDEX2026}

@article{Valeshabadi:2026ckt,
    author = "Kord Valeshabadi, R. and Rezaie, S. and Azizi, K.",
    title = "{Numerical study of MRW-type unintegrated double parton distribution functions from non-factorized DPDFs}",
    eprint = "2605.15853",
    archivePrefix = "arXiv",
    primaryClass = "hep-ph",
    doi = "10.1140/epjc/s10052-026-16264-0",
    journal = "Eur. Phys. J. C",
    volume = "86",
    number = "8",
    pages = "1002",
    year = "2026"
}

@article{Gaunt:2009re,
    author = "Gaunt, Jonathan R. and Stirling, W. James",
    title = "{Double Parton Distributions Incorporating Perturbative QCD Evolution and Momentum and Quark Number Sum Rules}",
    eprint = "0910.4347",
    archivePrefix = "arXiv",
    primaryClass = "hep-ph",
    reportNumber = "CAVENDISH-HEP-09-20",
    doi = "10.1007/JHEP03(2010)005",
    journal = "JHEP",
    volume = "03",
    pages = "005",
    year = "2010"
}

@article{Ceccopieri:2010kg,
    author = "Ceccopieri, Federico Alberto",
    title = "{An update on the evolution of double parton distributions}",
    eprint = "1011.6586",
    archivePrefix = "arXiv",
    primaryClass = "hep-ph",
    doi = "10.1016/j.physletb.2011.02.047",
    journal = "Phys. Lett. B",
    volume = "697",
    pages = "482--487",
    year = "2011"
}

@article{MRW,
    author = "Martin, A. D. and Ryskin, M. G. and Watt, G.",
title = "{NLO prescription for unintegrated parton distributions}",
eprint = "0909.5529",
archivePrefix = "arXiv",
primaryClass = "hep-ph",
reportNumber = "IPPP-09-78, DCPT-09-156",
doi = "10.1140/epjc/s10052-010-1242-5",
journal = "Eur. Phys. J. C",
volume = "66",
pages = "163--172",
year = "2010"
}

@article{Golec-Biernat:2016vbt,
    author = "Golec-Biernat, K. and Stasto, A. M.",
    title = "{Unintegrated double parton distributions}",
    eprint = "1611.02033",
    archivePrefix = "arXiv",
    primaryClass = "hep-ph",
    doi = "10.1103/PhysRevD.95.034033",
    journal = "Phys. Rev. D",
    volume = "95",
    number = "3",
    pages = "034033",
    year = "2017"
}

@article{Guiot:2022psv,
	author = "Guiot, Benjamin",
	title = "{Normalization of unintegrated parton densities}",
	eprint = "2205.02873",
	archivePrefix = "arXiv",
	primaryClass = "hep-ph",
	doi = "10.1103/PhysRevD.107.014015",
	journal = "Phys. Rev. D",
	volume = "107",
	number = "1",
	pages = "014015",
	year = "2023",
	note = "[Erratum: Phys.Rev.D 110, 039901 (2024)]"
}

@article{MSTW,
    author = "Martin, A. D. and Stirling, W. J. and Thorne, R. S. and Watt, G.",
    title = "{Parton distributions for the LHC}",
    eprint = "0901.0002",
    archivePrefix = "arXiv",
    primaryClass = "hep-ph",
    reportNumber = "IPPP-08-95, DCPT-08-190, CAVENDISH-HEP-08-16",
    doi = "10.1140/epjc/s10052-009-1072-5",
    journal = "Eur. Phys. J. C",
    volume = "63",
    pages = "189--285",
    year = "2009"
}

@article{Valeshabadi2026,
    author = "Kord Valeshabadi, R. and Rezaie, S.",
    title = "{PDFxTMDLib: A High-Performance C++ Library for Collinear and Transverse Momentum-Dependent Parton Distribution Functions}",
    eprint = "2412.16680",
    archivePrefix = "arXiv",
    primaryClass = "hep-ph",
    doi = "10.1007/s41781-025-00157-0",
    journal = "EPJ Res. Infrastruct.",
    volume = "10",
    number = "1",
    pages = "3",
    year = "2026"
}
\endgroup
\end{document}